# Dirac points with giant spin-orbit splitting in the electronic structure of two-dimensional transition-metal carbides

H. Fashandi, [1*†] V. Ivády, [1,2*] P. Eklund,[1] A. Lloyd. Spetz,[1] M. I. Katsnelson,[3,4] and I. A. Abrikosov[1,5]

[1] Department of Physics, Chemistry, and Biology, Linköping University, SE-581 83 Linköping, Sweden

[2] Wigner Research Centre for Physics, Hungarian Academy of Sciences, PO Box 49, H-1525, Budapest, Hungary

[3] Radboud University, Institute for Molecules and Materials, Heyendaalseweg 135, 6525AJ Nijmegen, The Netherlands

[4] Dept. of Theoretical Physics and Applied Mathematics, Ural Federal University, Mira str. 19, Ekaterinburg, 620002, Russia

[5] Materials Modeling and Development Laboratory, National University of Science and Technology 'MISIS', 119049 Moscow, Russia

KEYWORDS: Cone-point, MAX phase, MXene, Dirac fermion, Spin-orbit splitting

# ABSTRACT

Two-dimensional (2D) materials, especially their most prominent member, graphene, have greatly influenced many scientific areas. Moreover, they have become a base for investigating the relativistic properties of condensed matter within the emerging field of "Dirac physics". This has ignited an intense search for new materials where charge carriers behave as massless or massive Dirac fermions. Here, by the use of density functional theory and symmetry analysis, we theoretically show the existence of Dirac electrons in a series of 2D transition-metal carbides, known as MXenes. They possess eight conical crossings in the 1$^{st}$ Brillouin zone with giant spin-orbit splitting. Our findings indicate that the 2D band structure of MXenes is protected against external perturbations and preserved even in multilayer phases. These results, together with the broad possibilities to engineer the properties of these phases, make them a potential candidate for studying novel Dirac-physics-based applications.

# I. INTRODUCTION

The discovery of graphene [1] followed by the demonstration that charge carriers in graphene are massless Dirac fermions [2,3] has established a new and profound relation between materials science and fundamental physics. Many quantum relativistic phenomena such as Klein tunneling, Zitterbewegung, relativistic quantum collapse at supercritical charges, etc. have been studied using graphene as a model system [4]. "Dirac physics" turns out to be of crucial importance for many other materials such as topological insulators [5,6], and provided a base for a wide scope of studies, from two-dimensional (2D) material tailoring [7–10], to theoretical and experimental investigations of various condensed-matter systems with Dirac fermions [9–11]. Along this pathway, the investigation of novel 2D materials that possess graphene-like band structure, being generally a rare feature [12], while exhibiting various possibilities to tune their properties, robustness against external perturbations, or extended three-dimensional limits [13] potentially provide new directions for 2D technologies exploiting the interesting properties of Dirac fermions [14]. Additionally, the observation of substantial spin-orbit coupling in novel 2D Dirac materials could give rise to quantum spin Hall effect [5], being of great importance from both fundamental and applied points of view [15].

Advances in exfoliation methods have made layered materials a common source for 2D phases [16]. In this context, selective chemical etching of a series of $M_{n+1}AX_n$ phases [17] resulted in the synthesis of an emerging class of transition-metal-carbide-based 2D materials called MXene [18,19]. In the $M_{n+1}AX_n$ notation, M is an early transition metal (TM), A is an element mostly from groups 12-16, X is C or N, and n can be 1, 2, or 3. The name 'MXene' is defined to emphasize the selective etching of the A element from the $M_{n+1}AX_n$ phases as well as its similarity to graphene.

MXenes are 2D transition-metal carbide structures with hexagonal lattice, which can be prepared both in loosely packed multilayer and free-standing few-layer forms as well as in thin films [18–22]. Few-layer MXenes exhibit planar or curved configurations [18] and are transparent to visible light [20,22]. The surfaces of bare MXene sheets are expected to be chemically active. Thus, they are terminated by different atoms or chemical groups present in the synthesis process, which are fluorine (F), oxygen (O), and hydroxyl (OH) [18]. Using first principles simulations, methoxy-terminated MXenes were also investigated and predicted to be stable [23]. Recently, polymer functionalization of MXene surfaces was experimentally reported [24]. For multilayer MXenes, spontaneous intercalation of organic molecules and alkali ions has been reported by a number of experiments [20,22–24]. Khazaei *et al*. [25] theoretically examined a broad range of single-sheet MXene phases and showed that MXene phases with different surface terminations may possess electronic characteristics ranging from metallic to semi-conducting. These findings were also supported by other theoretical studies [18,26,27]. For some phases, magnetism [25,28] and large Seebeck coefficient and high thermoelectric power factor were also theoretically predicted [25,29]. Significant achievements have been demonstrated in the field of electrochemical energy storage and lithium ion batteries [30–34]. Moreover, there are several theoretical and experimental reports on other potential applications of MXene phases, e.g., sensing and detecting applications [35], utilization as chemical catalysts [23,36], applications as transparent conductors in solar cells [22], optical nano-devices [37], together with other reports [24,25,28].

However, less attention has been paid to studies of relativistic phenomena in MXenes (see Ref. [38] for a review). If existing, they would be of great interest, firstly, due to the presence of transition metals and the consequent attributes of the d-orbitals, secondly, the fact that there is a wide range of different MXenes which offers the opportunity to introduce a group of Dirac materials instead of one single phase, and thirdly, the presence of different termination elements/groups that can potentially cause additional influences on any electronic features, all being dissimilar to graphene. Therefore, we have carried out a systematic theoretical search, based on first principles calculations and group theory considerations, to explore the potential of MXenes for studying relativistic phenomena. We report the theoretical prediction of Dirac points at the energy bands near the Fermi level in a series of metastable MXenes. We examine the observed conical features of the scalar-relativistic band structure and study the effects of spin-orbit coupling and inter-laminar interactions on the Dirac points. Our results present MXenes as a promising model system for studying Dirac fermions.

The rest of the article is organized as follows. In Sec. II, we describe the first principle methods used in our study. In Sec. III the results on single sheet MXene phases, exhibiting conical features in their band structure which we refer to as Dirac MXenes, are presented, while in Sec. IV, the properties of a multilayer Dirac MXene phase are investigated. Finally, in Sec. V we summarize our findings.

## II.  METHODOLOGY

Our study is based on first-principles density functional theory (DFT) calculations. To describe the valence and core states, plane wave basis set and projector augmented wave method (PAW) [39,40] were used as implemented in the Vienna *ab initio* Simulation Package (VASP) [41]. We applied PBE parameterization of the generalized gradient approximation of the exchange and correlation energy functional [42]. Convergent sampling of the Brillouin-zone was achieved by a Γ-centered 15×15×1 grid. For the structural parametrization, $1\times10^{-3}$ eV/ Å force criteria was used. In order to model a single sheet of MXene we ensured at least 30 Å vacuum between the periodically repeated sheets. The relativistic spin-orbit coupling effects were also included in our simulations regarding band structure and density of states. [43] We considered Van der Waals interactions using the DFT-D2 approach of Grimme [44]. For the HSE06 calculations [45,46], we used the relaxed structure of the PBE calculation and 12×12×1 Γ-centered sampling of the BZ without including the Spin-orbit interaction. Phonon frequency calculations were done using Phonopy code [47] coupled with VASP.

## III. RESULTS AND DISCUSSIONS

### A. CRYSTAL STRUCTURES

We studied MXene phases based on the group 3-5 transition metals Sc, Ti, Zr, Hf, V, Nb, and Ta, described by $M_{n+1}C_n$ notation with n=1, 2, and 3. The choice of these elements is made from the definition of MAX phases [17]. Based on experimental results [19], we assumed the MXene sheets to be fully terminated by the chemical species present in the exfoliation process like O, OH, and F. The terminating species can be placed on three different sites, labeled A, B and T, shown in FIG. 1(a) for $M_2C$ stoichiometry. Sites A, B, and T refer to the hollow sites between three metal atoms, top of the carbon atoms, and top of the metal atoms, respectively. According to the previous first-principles calculations site T relaxes to one of the other configurations [25].

Considering full termination on both sides of the MXene sheets, three different possible configurations can be realized: AA, AB, and BB. In these notations, each letter shows the termination model on each side of the sheet. We assume AB configuration not to be stable in planar form, since our results show that the lattice constants of MXene sheets with pure AA termination model are ~2% larger than those of the corresponding BB models. This assumption is also supported by the phonon frequency dispersion analysis which shows negative frequencies for this configuration [48].

According to earlier theoretical reports, termination model AA is the most stable configuration for a wide range of terminated MXene phases [25,27]. However, the stability of the energetically less favorable configuration BB is less investigated. Therefore, we probed the potential energy surface with respect to the position of the termination atoms on the surface of single sheet non-terminated MXene. FIG.1 (b) shows the total energy values of $Ti_2C-F_2$ vs. the position of the terminating atoms along a path, marked by 'Path' in FIG.1 (a). The variation of the total energy along this path indicates that the AA configuration is the global minimum and BB is a local minimum, representing a metastable configuration. Moreover, FIG.1 (c) illustrates the phonon frequency dispersion of the same phase in BB configuration ($Ti_2C-F_2$-(BB)) showing all the frequencies to be positive, which is a feature of a dynamically stable phase. These result and the wide possibility of engineering different MXene phases with a wide range of M- and X-elements as well as different synthesis processes suggest that the BB configuration may be realized in experiment (see [48] for formation energy differences between the AA and the BB configurations of some MXenes).

From the symmetry properties point of view, all of these few-layer atomic sheet MXenes, ($M_{n+1}X_n$, n=1, 2, and 3), exhibit inversion symmetry, but lack in-plane reflection symmetry. The point group of theses sheets is $D_{3d} = D_3 \otimes i$, which is a subgroup of point group $D_{6h}$ of graphene. The addition of atomic termination layers on both sides of a $M_2X$ sheet (such as F or O) results in *c*(*abc*)*a* and *b*(*abc*)*b* stacking sequences along the high-symmetry axis of the 2D hexagonal lattice for the stable AA and for the metastable BB configurations, respectively (FIG.1 (a)). Our calculations for the important case of OH termination, showed that the favorable site of the hydrogen atom is on top of the oxygen atom (same in-plane coordinates), as a result, this structure is symmetrically equivalent to a mono-atom termination case. Importantly, all of the configurations mentioned above possess $D_{3d}$ point group symmetry, thus these terminated layers preserve inversion symmetry. The previous arguments can be easily generalized for terminated thicker $M_{n+1}C_n$ layers, with $n > 1$.

An important consequence of the $D_{3d}$ symmetry is that the presence of the inversion symmetry allows the appearance of accidental conical points in every general point of the Brillouin zone (BZ), as well as essential conical points in the high symmetry points of the BZ [49]. However, it does not necessarily mean the presence of Dirac-points at the Fermi-level that strongly depends on the interactions of the atomic orbitals forming the energy bands. *Ab initio* techniques can describe the energy bands properly and thus, can be utilized to investigate any potential appearance of Dirac points in the band structures of the numerous MXene phases.

### B. ELECTRONIC STRUCTURE

In this section, we study the properties of a class of MXene phases, where our first-principles calculations predict Dirac cones near the Fermi-level. We examine the scalar-relativistic and relativistic band structure, the density of states, and assess the phases that exhibit conical features in their band structure.

1. Scalar-relativistic approach

The calculated electronic band structure of one of the considered MXene phases, $Ti_2CF_2$-(BB), is presented in FIG. 2(a). The band structure is plotted along the high-symmetry lines of the hexagonal Brillouin zone (Γ-M-K-Γ). Without accounting for the spin-orbit interaction, the band structure shows conical band crossing at two points, one at the high-symmetry K-point and the other at a low-symmetry point between the K- and the Γ-points, which we mark as the F-point. In the 1$^{st}$ Brillouin zone, the two touching bands create six contact points at the K-points and another group of six at the F-points (FIG. 2(b)). While the former group is shared between 3 neighboring BZs and thus possesses the weight of 1/3, in total this system demonstrates 8 Dirac points.

These findings are independent of the choice of scalar-relativistic first principles method. To prove this, we compare the result of PBE and HSE06 calculations. The latter method is more sophisticated and computationally more demanding due to the inclusion of a fraction of the exact exchange interaction. The HSE06 bands at the Fermi-level together with the PBE ones are depicted in FIG.3. As expected, the HSE06 bands exhibit the same two-cone structure as the PBE bands, however, the energy of the Dirac-point at the K-point as well as the energy and the place of the accidental Dirac-point at the F-point are shifted. Note that the observed difference does not affect our conclusions and qualitative statements.

The energy bands in the vicinity of the Γ-point for $Ti_2CF_2$-(BB) are rather flat (see FIG. 2(a)) resulting in a high density of the states (DOS) near the Fermi level. FIG.4 shows the total DOS of Ti$_2$CF$_2$-(BB) together with the partial densities projected onto $d_{xy} + d_{x^2-y^2}$ and $d_{z^2}$ orbitals, calculated by PBE functional. The partial DOS plots show that at the Fermi level, where the cone points are located, both the occupied and unoccupied peaks in the DOS mostly originate from the $d_z^2$ orbital of Ti.

We also found such conical points in the vicinity of the Fermi level in the electronic band structure of many other metastable MXene phases, such as M$_2$C-T$_2$-(BB), where M = Ti, Zr, and Hf for T = F and OH, and M$_2$C-O$_2$-(BB) where M = V, Nb or Ta (see FIG.5). Beside the structure, the common feature in all these phases is that the conduction band is filled with the same number of electrons. One valence electron per TM atom is approximately enough to occupy the lowest lying conduction band ($d_I$). Therefore, the Fermi-level is pinned at the top of $d_I$ band. Note, however, that the position of the two Dirac points with respect to the Fermi-level as well as the position of the F-point is changing among these different MXene phases. The stable AA configurations as well as thicker phases (n > 1), generally do not exhibit conical features at the Fermi level. Similar features for DOS are seen in M$_2$CF$_2$-(BB) MXene phases with M= Zr and Hf, [48] making it possible to realize a mixed system of Dirac-fermions and ordinary conducting electrons in all the phases discussed.

## 2. GROUP THEORY ANALYSIS

In the following, by applying group theory considerations, we analyze the energy bands exhibiting the multiple conical features in the scalar-relativistic band structure of Ti$_2$C-F$_2$-(BB). Our statements, however, are also valid for MXene phases of M = Zr and Hf with OH or F termination. For the case of M = V, Nb, and Ta with O termination, only the statements related to the bands near the K point of the BZ are valid.

As the symmetry of the structure is $D_{3d}$, the point group of the wave-vector at the high-symmetry point Γ, K, and M is $D_{3d}$, $D_3$, and $C_{2h}$, respectively. Furthermore, the point group along the high symmetry lines Γ-K, K-M, and M-Γ is $C_2$, $C_2$ and $C_{1v}$ respectively, and it is $C_1$ in every other non-symmetrical points of the hexagonal BZ ( see Table.I ) [50]. 2D irreducible representation belongs only to the $D_{3d}$ and $D_3$ point group of the Γ- and the K-points, respectively, while every other point groups possess only one-dimensional irreducible representations. Therefore, two-fold degeneracy, claimed by the symmetry, can only appear at the Γ- and the K-points of the BZ. As a consequence, the observed Dirac point at the K-point is an essential degeneracy of the bands forced by symmetry, similar to graphene and silicene, while the one observed at the F-point is an accidental degeneracy, which appears due to the crossing of the energy bands.

Next, we provide the symmetry classification of the two energy bands ($d_I$ and $d_{II}$), which form the conical features at the F- and the K-points of the BZ. By using the character tables of the point groups of the high symmetry wave-vectors and by analyzing the symmetry properties of the wave functions of the states, we could identify the irreducible representations of the point groups of the considered bands. The results are given in Table I and depicted in FIG.6. Note that the irreducible representations of the high symmetry points satisfy the compatibility relations.

By looking at the bands along the high symmetry lines of the BZ, one can realize that $d_I$ and $d_{II}$ bands have different representations A and B along the lines Γ-K and K-M, while they have the same representations A' along the line M-Γ. As bands of different symmetries cannot interact and mix with each other, they can exhibit band crossing, (see $d_I$ and $d_{II}$ bands along line Γ-K). However, when bands of the same symmetry approach each other, they can mix and thus they exhibit avoided crossing or anti-crossing, as it is observed along line M-Γ. Furthermore, since the symmetry of a general wave vector k is $C_1$, which has only one irreducible representation giving all the bands the same representation, such avoided crossing of the energy band occurs regularly for every general non-symmetric wave vector. Therefore, as crossing is only allowed along the lines Γ-K and K-M, the crossing point is necessarily a single point in the reciprocal space and the dispersion relation is linear at the vicinity of the crossing point. From these considerations, it can be seen that when two bands possessing the symmetry properties of $d_I$ and $d_{II}$ cross each other along line Γ-K or K-M they necessarily form a Dirac point in the irreducible BZ. Also, due to symmetry reasons they produce six cones at the same time in the symmetrical BZ.

Next, we investigate why these bands cross at all. To answer, we compare the symmetry classification of the bands $d_I$ and $d_{II}$ with those of the bands π and π* of graphene. The bands around the K-point are nearly identical for graphene and MXene. In both cases, starting from the high symmetry K-point, the band of representation B and A decreases and increases in energy, respectively, by following the line Γ-K, while the opposite trend can be observed along the line K-M. On the other hand, the band structure is different for graphene and MXene at the Γ-point. For graphene, the π-band of representation $B_2$ ends in the irreducible representation $A_{2u}$ at the Γ-point, while π*-band of representation $A_2$ ends in the irreducible representation $B_{2g}$. An important difference between the irreducible representations $A_{2u}$ and $B_{2g}$ is that the former is symmetric under $C_2$, $C_3$, and $C_6$ rotation operations, while the latter is less symmetric and the eigenstates change sign upon $C_2$, $C_{2-1}$ and $C_6$, $C_{6-1}$ rotations. Such symmetry properties require the presence of three nodal planes perpendicular to the plane of graphene and thus, adumbrate higher energy of the eigenstate of representation $B_{2g}$. As a result, starting from the double degenerate state of the K-point, the π-band of representation $B_2$ monotonously decreases in energy to reach the highly symmetric low energy representation $A_{2u}$ at point Γ, while the π*-band monotonously increases in energy to reach the less symmetric high energy representation $B_{2g}$ at the Γ-point. For the considered MXene phases, one has to examine the $A_{1g}$ and $E_g$ representations of $D_{3d}$ at the Γ-point. Since $A_{1g}$ is fully symmetric, one can expect lower energy for the eigenstate of representation $A_{1g}$ while being higher for $E_g$. Therefore, here an opposite trend can be observed as for graphene: The band of representation B starts to decrease in energy as leaving the K-point but ends in the less symmetric higher energy representation $E_g$ at the Γ-point, while the band of representation A starts to increase but finally result in a fully symmetric low energy representation $A_{1g}$. Such behavior necessarily results in a crossing of the bands, resulting in an accidental Dirac-point in the irreducible BZ close to the Γ-point.

In Table I, one can also see the d-orbital decomposition of the charge density of the eigenstates of the different irreducible representations. At the K-point, the two eigenstates of the representation E are separately localized on the two TM layers of the MXene sheet and build up from the in-plane $d_{xy}$ and $d_{x2-y2}$ orbitals. At the Γ-point, the fully symmetric $A_{1g}$ state builds up from the $d_{z2}$ orbitals perpendicular to the plane of the MXene. Eigenstates of representation $E_g$ are also made of in-plane $d_{xy}$ and $d_{x2-y2}$ orbitals.

Next, we shortly discuss the differences between the electronic structure of configurations AA and BB of terminated MXene sheets, which causes the disappearance of the Dirac points for the former case. By comparing the two band structures one can realize that band $d_I$ is shifted upward in configuration AA, thus the $A_{1g}$ and $E_g$ states change their order at the Γ-point, (see FIG.7). A crossing feature, however, appears 1 eV above the Fermi-level at point K. The different behavior of bands $d_I$ and $d_{II}$ upon the change of the stacking of the termination layers can be understood by looking at the d-orbital decomposition of the bands. As can be seen in Table I, the charge density of band $d_I$ contains substantial contribution from the $d_z^2$ orbitals. In the case of BB configuration, the termination layers are above and below the carbon atoms, in b(abc)b stacking, and only slightly affect the $d_z^2$ orbitals of the TM atoms. On the other hand in the AA configuration the terminating atoms are above and below the TM atoms, in c(abc)a stacking, (see Fig.1(a)). The presence of the termination atoms can represent an additional potential constraint on the $d_z^2$ orbitals of TM atoms that can raise the energy of the band of high $d_z^2$ contribution. Here, we would like to note that $d_I$ state is not a bonding state between the TM atoms and the termination. On the other hand, the $d_{II}$ band is mainly built up from in-plane d-orbitals, see TableI, which are much less affected by the termination layers. This explains the exchange of the $A_{1g}$ and $E_g$ orbitals at the Γ-point and the shifting of the Dirac-point at the K-point.

As a final remark, one should keep in mind that the discussion above is intended to explain the presence or absence of conical features in the scalar-relativistic band structure of MXene sheets; however, as the spin-orbit interaction is taken into account, it lifts the degeneracy of the conical points and opens a substantial gap.

## 3. CONSIDERATION OF THE SPIN-ORBIT COUPLING

Spin-orbit interaction splits the bands at the contact points (see FIG.8). This splitting is to some extent similar to what was predicted by Kane and Mele for graphene [5]. The splitting magnitude is many orders of magnitude larger than in flat graphene (where it is of the order of μeV [4,6]), being at the order of 10 meV (1000 K) for Zr- or Hf-based $M_2CF_2$-(BB) phases (see Table II.). This makes the new potential MXene phases interesting for studies of spin quantum Hall effect, in addition to other known 2D topological insulators [6]. The magnitude of splitting caused by spin-orbit coupling effect at the F-point is approximately 20% of that of the K-point for $M_2CF_2$-(BB), M=Ti, Zr, and Hf. In addition, the splitting increases with the row-index of the M element.

## IV. MULTILAYERED MXENE

Inter-laminar MXene-MXene interactions are also of importance, since experimental synthesis of MXenes is based on exfoliation of bulk or thin-film $M_{n+1}AX_n$ phases, resulting in MXene layers forming multi-layer powder or film. Recently, Halim *et al.* reported typical 2D-type electronic transport behavior for $Ti_3C_2$-based thin film MXene [22], indicating that the inter-laminar MXene-MXene interactions in bulk form may be weak enough to preserve certain electronic features of single-sheet MXene. To further understand this behavior, we depict the charge densities of termination related deep lying bands as well as the cone forming bands $d_I$ and $d_{II}$ at the Fermi-level for the case of $Ti_2CF_2$-(BB) in FIG. 9. In the latter case, the charge density is localized only on the Ti sub-layers, which is in accordance with our results presented above. On the other hand, the charge density of F-related bands correspond to atomic closed shell (F⁻)-like states, which can only weakly interact with the ambient. This outer layer of the MXene sheets can protect the conducting states close to the Fermi energy and preserve single-sheet properties.

In order to investigate the effect of MXene-MXene interactions on the band structure, we studied bulk $Ti_2CF_2$-(BB) system by including Van der Waals interactions. The primitive cell is a double-layer MXene sheet, which is periodically repeated in all directions (see FIG.10 (a)). The scalar-relativistic dispersion of the bands along the (Γ-M-K-Γ) path is illustrated in FIG. 10 (b). It shows quasi-degenerated bands resembling those of the single sheet MXene. FIG. 10 (c) and (d) show the band crossings at the K- and the Γ-points, respectively. As one can see, the existing interaction destroys the ideal conical points in both points and produces either splitting or touching of parabolic bands, the latter is similar to the bilayer graphene [4]. However, farther away from the contact area the band structure possesses a quasilinear dispersion relation. The splitting at the K-point is 10.8 meV between the highest and lowest bands. By including spin-orbit interaction in our bulk model, we obtained two quasi-double degenerated bands exhibiting 25.8 meV splitting at the K-point (see FIG. 10 (c)), which is close to that of the single sheet phase (see Table II). This result indicates that the effect of the spin-orbit interaction dominates over that of the inter-laminar MXene-MXene interactions at the K-point and thus, the bulk system may electronically behave as an ensemble of independent single MXene sheets.

V.   FINAL REMARKS AND SUMMARY

The presented demonstration of the conical points close to the Fermi energy makes the whole concept of "Dirac physics" relevant for this emerging family of 2D transition-metal carbides. The most important experimental manifestations at the temperatures or chemical potentials higher than the spin-orbit splitting may be existence of an analog of "minimal metallic conductivity" of the order of $e^2/h$ and relatively low sensitivity of electric properties to a smooth external disorder, due to Klein tunneling [4]. At high enough doping, when the Fermi energy is far enough from the conical point it will manifest itself in the appearance of the Berry phase π changing the sign of Shubnikov – de Haas oscillations of the conductivity in magnetic field [4]. Strong spin-orbit coupling makes the application of the Kane-Mele model [5] originally proposed for graphene with its very weak spin-orbit coupling of interest to this new system, though several issues, e.g. d-orbital character of the cone forming bands and the presence of the two sets of Dirac points, distinguish Dirac MXenes from graphene. In Zr- and Hf-based systems at the K-points, the gap may be higher than the equivalent room temperature energy. Existence of topologically protected edge states is another dramatic consequence. Of course, the absence of the global band gap (see FIG. 2.), hinders immediate application of Dirac MXenes in this direction and requires further investigations. Topological insulator states could be realized by appropriate engineering the properties of MXene sheets. For instance, as the two Dirac cones exhibit different properties and stability, a properly chosen symmetry breaking termination layer could open a substantial gap at the F-point, while keeping the cone at the K-point unchanged.

In summary, we demonstrated that the band structure of a class of MXene phases exhibits multiple Dirac cones and giant spin-orbit splitting. The intriguing features of the electronic structure are protected against external perturbations and can be preserved even in multilayer phases. The possible control over the properties of MXenes together with the experimental advances in synthesis of these materials should stimulate the experimental realization of the aforementioned properties. Our findings suggest numerous new applications for these recently synthesized 2D materials and provide them as new and promising target for studying Dirac physics. The reported protection mechanism, which characterizes these materials, may lead to the development of robust and applicable 2D technologies based on Dirac fermions.

ACKNOWLEDGMENTS

We acknowledge Elham Mozafari for help in running the simulations and discussions about the results, and Jens Eriksson and Mike Andersson for discussions and support. We acknowledge the support from VINN Excellence Center in Research and Innovation on Functional Nano-scale Materials (FUNMAT) by the Swedish Governmental Agency for Innovation Systems (VINNOVA) and the Swedish Foundation for Strategic Research through the Synergy Grant FUNCASE. The simulations were done using supercomputer resources provided by the Swedish national infrastructure for computing (SNIC) carried out at the National supercomputer center (NSC) and the Center for high performance computing (PDC). V.I. and I.A.A. acknowledge support from the Knut & Alice Wallenberg Foundation "Isotopic Control for Ultimate Materials Properties", the Swedish Research Council (VR) Grants No. 621-2011-4426 and 621-2011-4249, the Swedish Foundation for Strategic Research program SRL grant No. 10-002, and the Grant of the Ministry of Education and Science of the Russian Federation (grant No. 14.Y26.31.0005). P.E also acknowledges the Swedish Foundation for Strategic Research through the Future Research Leaders 5 program. M.I.K. acknowledges support from Fundamenteel Onderzoek der Materie (FOM) and De Nederlandse Organisatie voor Wetenschappelijk Onderzoek (NWO).

[*]These authors contributed equally to this work.

[†] E-mail: hosfa@ifm.liu.se

FIGURES:

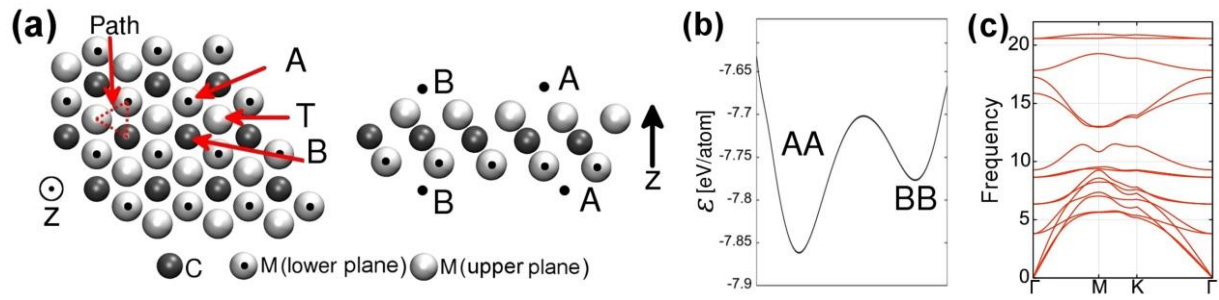

FIG.1. Structure of a MXene sheet. (a) Surface- and side-view of a non-terminated $M_2C$ MXene sheet showing C- and metal (M) atoms, different termination positions, and the path used for energy comparison of different termination models. (b) Total energy comparison between AA and BB models of $Ti_2CF_2$. (c) Phonon dispersion frequencies of $Ti_2CF_2$-(BB)

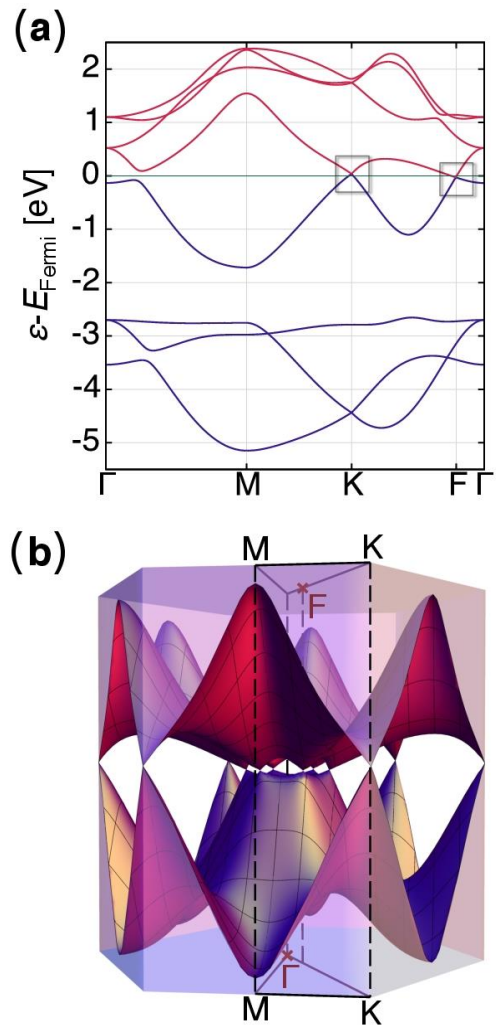

FIG.2. Band structure of $Ti_2CF_2$-(BB) showing Dirac points (a) Energy dispersion of the bands along the high symmetry lines of the two-dimensional Brillouin zone. The Fermi energy is chosen to be the zero value of the energy scale. (b) 3D plot of the cone forming bands at the Fermi level over the 2D symmetrical Brillouin zone.

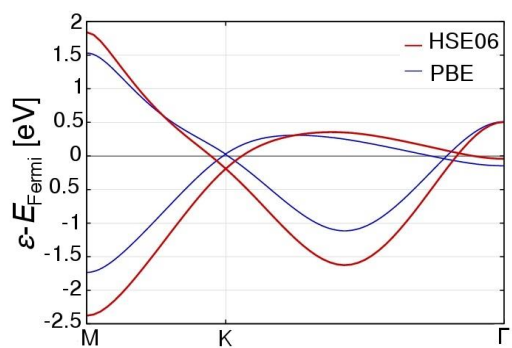

FIG.3 Comparison of HSE06 and PBE scalar-relativistic band structure at the Fermi energy.

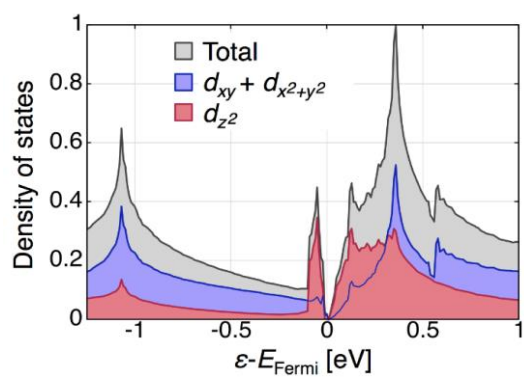

FIG.4. Density of states (DOS) of the Dirac MXene $Ti_2CF_2$-(BB). Total and partial DOS projected onto $d_{xy} + d_{x^2-y^2}$ and $d_{z^2}$ orbitals, respectively, are depicted as a function of energy, relative to the Fermi level.

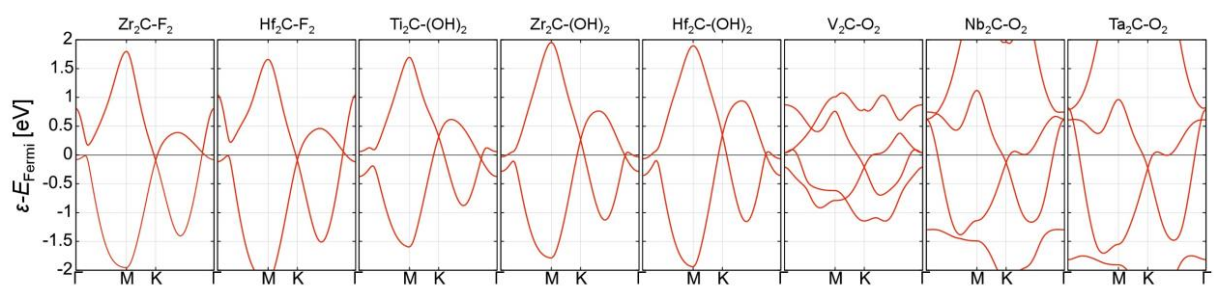

FIG.5. Band structure plots of a series of MXene phases showing the closest bands to the Fermi level.

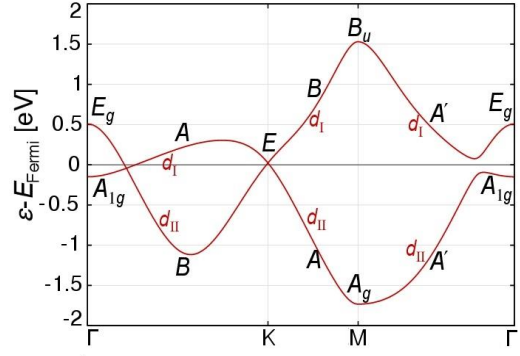

FIG.6. Band structure of Ti$_2$C-F$_2$ (BB) MXene sheet near the Fermi level. The point group representations are also indicated.

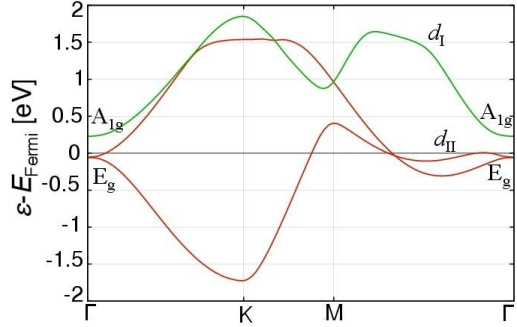

FIG.7. Band structure of Ti$_2$C-F$_2$-(AA) MXene sheet around the Fermi level. The symmetries of the states at the Γ-point of the BZ are given. The states show the opposite order as for the BB configuration.

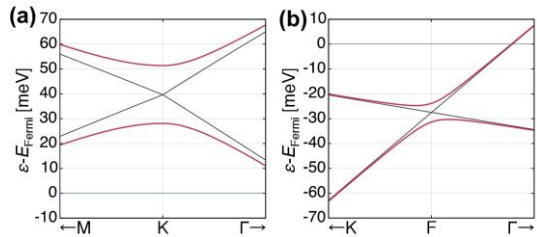

FIG.8. Spin-orbit splitting of the Dirac points for Ti$_2$CF$_2$-(BB). Energy dispersions of the bands near (Left) the K- and (Right) the F-point of the Brillouin zone. Thin black lines and thick red lines show the results of scalar relativistic and spin-orbit calculations, respectively. The Fermi energy is chosen to be the zero value of the energy scale.

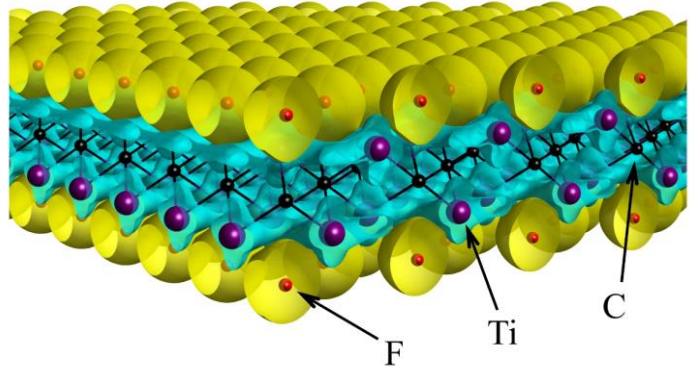

FIG.9 Partial charge density of the cone forming bands (blue lobes) and the termination-atoms-related bands (yellow lobes) of the Dirac MXene $Ti_2CF_2$-(BB) are depicted along with the corresponding structure.

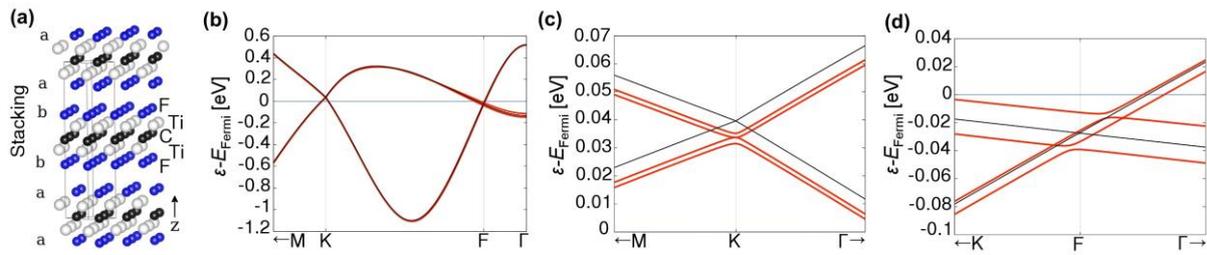

FIG.10. (a) Cell of bulk $Ti_2CF_2$-(BB) MXene phase, showing the primitive cell and the stacking sequence of the F atoms.+ (b) Energy dispersion of the two bands at the Fermi level of bulk $Ti_2CF_2$-(BB) MXene. The framed areas are depicted separately in (c) and (d) showing the spin orbit splitting of the bands. Thin black lines and thick red lines show the results of scalar relativistic and spin-orbit calculations, respectively.

TABLES:

Table I. Symmetry classification of the two energy bands forming the conical points. The case of graphene [50] is presented for comparison.

| Point, line | | $\Gamma$ | $\Gamma$-K | K | K-M | M | M-$\Gamma$ |
|---|---|---|---|---|---|---|---|
| Single sheet MXene: $M_2C$-$T_2$, BB configuration (M = Ti, Zr, Hf; T = F, OH) | | | | | | | |
| Symmetry | | $D_{3d}$ | $C_2$ | $D_3$ | $C_2$ | $C_{2h}$ | $C_{1v}$ |
| Band $d_I$ | Rep. | $A_{1g}$ | $A$ | $E$ | $B$ | $B_u$ | $A'$ |
| | Charge density | $1/2(d_{z2}^{(1)}+d_{z2}^{(2)})$ | | $1/2(d_{xy}^{(1)}+d_{x2-y2}^{(1)})$ | | $1/2(d_{z2}^{(1)}+d_{z2}^{(2)})$ | |
| Band $d_{II}$ | Rep. | $E_g$ | $B$ | $E$ | $A$ | $A_g$ | $A'$ |
| | Charge density | $3/8(d_{xy}^{(1)}+d_{xy}^{(2)}) + 1/8(d_{x2-y2}^{(1)}+d_{x2-y2}^{(2)})$ | | $1/2(d_{xy}^{(2)}+d_{x2-y2}^{(2)})$ | | $1/4(d_{xy}^{(1)}+d_{xy}^{(2)}+d_{z2}^{(1)}+d_{z2}^{(2)})$ | |
| Graphene | | | | | | | |
| Symmetry | | $D_{6h}$ | $C_{2v}$ | $D_{3h}$ | $C_{2v}$ | $D_{2h}$ | $C_{2v}$ |
| $\pi$ | Rep. | $A_{2u}$ | $B_2$ | $E''$ | $A_2$ | $B_{2g}$ | $B_2$ |
| $\pi^*$ | Rep. | $B_{2g}$ | $A_2$ | $E''$ | $B_2$ | $B_{1u}$ | $B_2$ |

Table II. The splitting in the band crossing of $M_2CF_2$-(BB) MXene phases with M=Ti, Zr, and Hf, at the K- and the F- points caused by spin-orbit coupling effect.

| M-element | Energy splitting at K-point (meV) | Energy splitting at F-point (meV) |
|---|---|---|
| Ti | 23 | 5 |
| Zr | 83 | 19 |
| Hf | 86 | 20 |

# Dirac points with giant spin-orbit splitting in the electronic structure of two-dimensional transition-metal carbides

*H. Fashandi,* [1] *V. Ivády,* [1,2] *P. Eklund,*[1] *A. Lloyd Spetz,*[1] *M. I. Katsnelson,*[3,4] *and I. A. Abrikosov*[1,5]

[1] Department of Physics, Chemistry, and Biology, Linköping University, SE-581 83 Linköping, Sweden

[2] Wigner Research Centre for Physics, Hungarian Academy of Sciences, PO Box 49, H-1525, Budapest, Hungary

[3] Radboud University, Institute for Molecules and Materials, Heyendaalseweg 135, 6525AJ Nijmegen, The Netherlands

[4] Dept. of Theoretical Physics and Applied Mathematics, Ural Federal University, Mira str. 19, Ekaterinburg, 620002, Russia

[5] Materials Modeling and Development Laboratory, National University of Science and Technology 'MISIS', 119049 Moscow, Russi

# I. FORMATION ENERGY OF CONFIGURATION AA AND BB

Here we would like to give some additional computational details on the study of the energy difference between AA and BB models, presented in the main text. To study the relative stability of AA and BB configurations from the total energy point of view, we fixed the in-plane coordinates of the termination atoms in the primitive cell, but allowed them to relax in the perpendicular direction to the sheet (Z-direction). Prior to this simulation we relaxed the non-terminated sheet. The path on which we moved the terminating atoms is illustrated in FIG. 1 (a) in the main text, marked by 'Path'. The path totally consists of 33 steps, starts from TT to AA, BB and ends at TT configuration from where it starts.

As we mentioned in the main text the AA configuration is stable while the BB configuration is metastable. On the other hand, the formation energy difference between the two configurations varies among the different MXene phases. In Supplementary Table 1, one finds the formation energy differences for the important cases. It can be seen that the OH-terminated layers exhibits lower energy differences than F terminated ones, furthermore, it is also clear that the energy difference decreases by moving from Ti toward Hf based MXenes sheets. The replacement of the F termination to Cl or Br termination could also lower the difference.

The largest and smallest energy differences we obtained are 109 and 19 meV/atom for $Ti_2C-F_2$ and $Zr_2C-(OH)_2$ MXenes, respectively (Table 1). The reduction of the formation energy difference can greatly affect the equilibrium concentration of the AA and BB phases, which should promote the experimental realization of the metastable configurations.

**Supplementary Table 1.** Formation energy difference of AA and BB configurations of potentially interesting MXene phases.

| Phase | $Ti_2C-F_2$ | $Zr_2C-F_2$ | $Hf_2C-F_2$ | $Ti_2C-(OH)_2$ | $Zr_2C-(OH)_2$ | $Hf_2C-(OH)_2$ |
|---|---|---|---|---|---|---|
| $H_{BB}-H_{AA}$ [meV/atom] | 109 | 66 | 60 | 47 | 19 | 20 |

## II. TOTAL AND PARTIAL DOS OF ZR- AND HF-BASED $M_2CF_2$-(BB) PHASES

Similar features to $Ti_2CF_2$-(BB) are seen in $M_2CF_2$-(BB) MXene phases with M= Zr and Hf

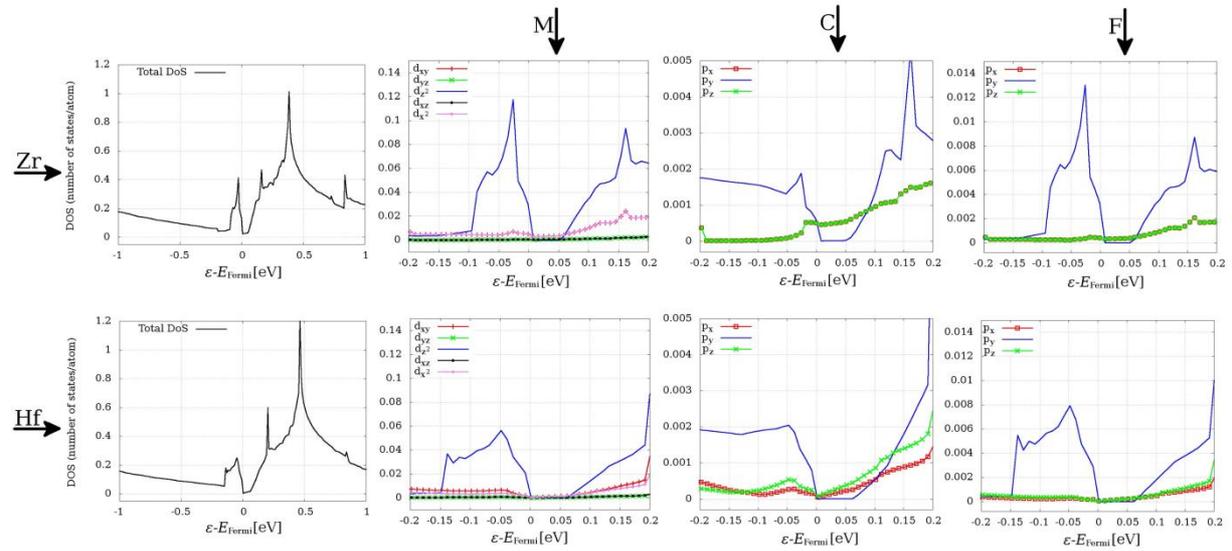

**Supplementary Figure 1.** Total and partial DOS. Different $M_2CF_2$-(BB) MXene phases with M= Zr and Hf.

## III. Phonon frequency dispersion of Ti$_2$CF$_2$-(AB)

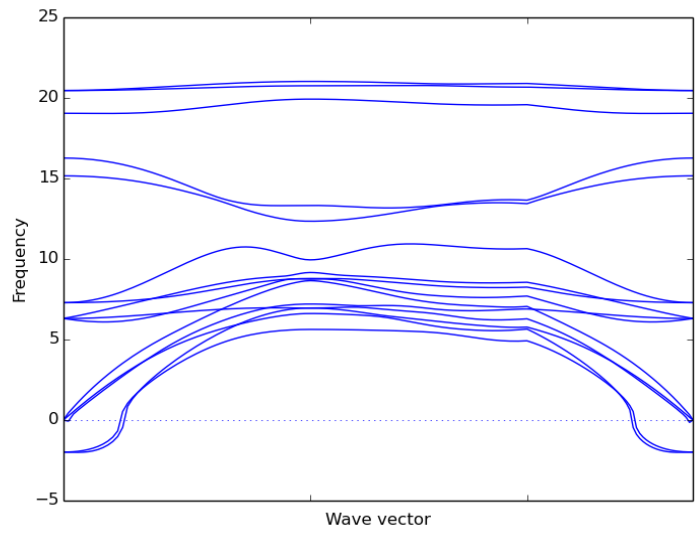

**Supplementary Figure 2.** Phonon frequency dispersion of Ti$_2$CF$_2$-(AB)